\documentclass[aps,prl,groupedaddress,twocolumn]{revtex4-2}

\usepackage{graphicx}
\usepackage[percent]{overpic}
\usepackage{braket}
\usepackage{color}
\usepackage{multirow}
\usepackage{booktabs}
\usepackage[table,xcdraw]{xcolor}
\usepackage{bm}
\usepackage{blindtext}
\usepackage{subfigure}
\usepackage{amsmath,amssymb}
\usepackage{subfig}
\usepackage{amsmath}
\usepackage[normalem]{ulem}

\newlength\myHeight 
\newlength\myWidth

\usepackage[overlay,absolute]{textpos}

\newcommand\PlaceText[3]{

\begin{textblock*}{10in}(#1,#2)

#3

\end{textblock*}

}

\begin{document}

\newcommand{\myTitle}{Simplified intensity- and phase-modulated transmitter for modulator-free decoy-state quantum key distribution} 

\newcommand{\myAuthors}{
    Y.~S.~Lo$^{1, 2}$,
	R.~I.~Woodward$^{1}$,
	N.~Walk$^{1}$,
	M.~Lucamarini$^{1,3}$,
	I.~De Marco$^1$,\\
	T. K.~Para\"{i}so$^1$,
	M. Pittaluga$^1$,
	T.~Roger$^1$,
	M.~Sanzaro$^1$,
	Z.~L.~Yuan$^1$,
	A.~J.~Shields$^1$
}

\newcommand{\myAffiliations}{
	$^1$Toshiba Europe Ltd, Cambridge, UK\\
	$^2$Quantum Science \& Technology Institute, University College London, UK\\
	$^3$Department of Physics and York Centre for Quantum Technologies, University of York, YO10 5DD York, UK \\
}

\newcommand{\myEmail}{yuen.lo@crl.toshiba.co.uk}

\newcommand{\eqn}[1]{\begin{eqnarray} \newline #1 \end{eqnarray}}
\newcommand{\ee}{&=&}
\newcommand{\ketn}[1]{\left |#1\right \rangle}
\newcommand{\nn}{\nonumber}
\newcommand{\hs}{\hspace{0.2cm}}
\newcommand{\bk}[1]{\left ( #1\right )}

\title{\myTitle}
\author{\myAuthors}
\affiliation{\myAffiliations}
\email[]{\myEmail}

\begin{abstract}

Quantum key distribution (QKD) allows secret key exchange between two users with unconditional security. For QKD to be widely deployed, low cost and compactness are crucial requirements alongside high performance. Currently, the majority of QKD systems demonstrated rely on bulk intensity and phase modulators to generate optical pulses with precisely defined amplitude and relative phase difference---i.e. to encode information as signal states and decoy states. However, these modulators are expensive and bulky, thereby limiting the compactness of QKD systems. Here, we present and experimentally demonstrate a novel optical transmitter design to overcome this disadvantage by generating intensity- and phase-tunable pulses at GHz clock speeds.
Our design removes the need for bulk modulators by employing directly modulated lasers, in combination with optical injection locking and coherent interference.
This scheme is therefore well suited to miniaturization and photonic integration and we implement a proof-of-principle QKD demonstration to highlight potential applications.

\end{abstract}


\maketitle

\PlaceText{12mm}{8mm}{APL Photonics \textbf{8}, 036111 (2023); https://doi.org/10.1063/5.0128445}

\section{Introduction}

Quantum key distribution (QKD) allows two parties to exchange secret keys with security guaranteed by the fundamental laws of physics \cite{ Gisin2002, bb84}. Driven by its potential, tremendous progress has been made in both theoretical and technological developments, such as satellite-based QKD \cite{Liao2017, satellite2}, QKD networks \cite{Sasaki2011,Stucki_2011, Dynes2019, Chen2021}, chip-based QKD \cite{Paraiso2019, Bunandar2018, Sibson2017} as well as the invention of novel protocols allowing higher secret key capacity \cite{Lucamarini2018, tf2, Pittaluga2021}.

In QKD protocols, time-bin encoding is commonly used \cite{Boaron2018, Yuan2018, Frohlich2013, Tang2016b} where the temporal modes of a time-bin qubit (early and late time bins) and the phase between them are used to encode the key bits. As practical single photon sources are not yet widely available, QKD systems typically employ lasers to generate weak coherent states to approximate the time-bin qubits. Since the photon number statistics of laser emission follows a Poisson distribution, the emitted pulses have a non-negligible probability of containing more than one photon, making laser-based QKD systems susceptible to a photon-number-splitting (PNS) attack \cite{Brassard2000}. Although it is still possible to obtain unconditional security, the signal flux has to be heavily attenuated in order to suppress multi-photon emission, thus giving a poor scaling of the secure key rate with transmission distance \cite{Gottesman2004}. Fortunately, this problem can be overcome by employing the decoy state method \cite{Hwang2003, Lo2005}: in addition to sending signal states, one also randomly sends a small number of states with reduced intensity, known as decoy states. A potential eavesdropper cannot distinguish between signal states and decoy states, thus any attempt to perform photon-number-dependent attacks can be detected from the measured photon statistics. As a result, with the decoy state method, single-photon bounds can be reliably estimated and therefore improve the scaling of secure key rate with distance significantly. 

Implementing a decoy-state QKD transmitter requires the ability to on-off modulate each time bin within a state, modulate the phase between time bins, as well as vary the intensity level to generate decoy states. To date, this has been achieved by placing intensity modulators after a light source to control the output intensity and phase modulators are also required in order to encode the phase information. Conventional intensity and phase modulators are based on LiNbO$_3$ crystals. While these modulators are widely available and offer high performance, they are expensive, bulky (centimeter-scale) and require high driving voltage (typically $>$ 4V) which often necessitates the addition of amplifiers. It is therefore highly beneficial to develop an alternative approach that can replace such modulators as it would significantly reduce the overall complexity, making QKD systems more compact and cost-effective.

Recently, Yuan \textit{et al} demonstrated an efficient scheme to perform direct phase modulation without the need for phase modulators \cite{Yuan2016}. Precise phase control is enabled by exploiting optical injection locking (OIL) and gain-switching techniques. Following this work, direct phase modulated laser transmitters for QKD have been studied more widely~\cite{Paraiso2021}, bringing the benefits of compact low-drive-voltage phase modulation for chip-based QKD \cite{Paraiso2019} as well as other emerging protocol such as measurement-device-independent QKD \cite{Woodward2021a}. More recently, the theoretical aspect of the direct phase modulation scheme has also been studied, verifying its favourable features in practical usage \cite{Shakhovoy2021}. While this scheme allows phase information to be directly encoded, it cannot be used to control the intensity of pulses for decoy state generation. Since a direct intensity modulation scheme is still missing, the use of bulk intensity modulators has been unavoidable.

In this work, we present a novel approach that can directly generate intensity \textit{and} phase modulated optical pulses.
Our scheme only requires two laser diodes and a passive asymmetric Mach-Zehnder interferometer (AMZI).
Such pulse source can generate all encoding states required for decoy-state QKD, thereby eliminating the need for external modulators and opening a new route for the development of compact, cost-effective and high-performance QKD systems.

\section{Direct generation of encoding states }

\begin{figure*}[ht!]
  \centering
  \subfigure[ ]{\includegraphics[height=151.766pt,width=269.806pt]{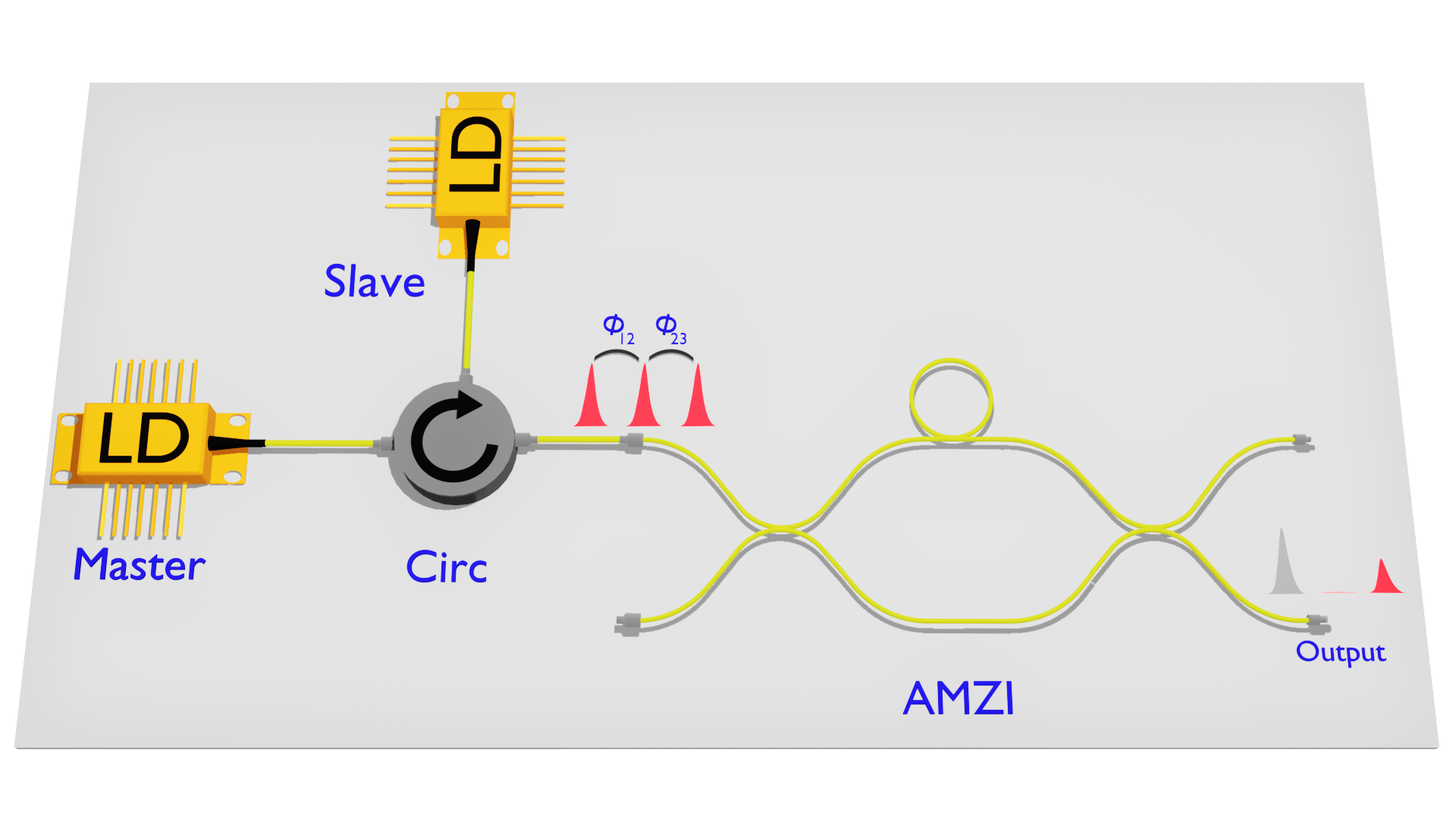}}\quad
  \subfigure[ ]{\includegraphics[height=186.627pt,width=205.012pt]{scheme_b1.pdf}}
\caption{Directly intensity and phase modulated transmitter scheme: (a) The design of the transmitter based on two directly modulated lasers in optical injection locking setup and an AMZI. (b) Schematic illustration of operating principle.}
\label{scheme}
\end{figure*}

Our scheme further extends direct phase modulation techniques~\cite{Yuan2016} by generating and interfering three intermediate pulses with carefully crafted relative phases, in order to accurately control both the relative phase and intensity of the final output pulses. The experimental setup is shown in Fig. \ref{scheme}a. The two laser diodes (referred to as `master' and `slave', following standard nomenclature) are connected in an OIL configuration. The master laser is gain-switched such that the laser produces long pulses (i.e. with high duty cycle) when it is driven above the threshold and switched off between the pulses. As a result, each pulse is produced with a random phase as they are seeded by spontaneous emission photons \cite{Yuan2014}. Subsequently, these pulses are injected through a circulator into the slave laser, which is gain-switched to produce three short pulses within each long master pulse [see Fig \ref{scheme}b (i)-(ii)]. Because the stimulated emission of the slave laser is seeded by the injected photons, the three slave pulses inherit the phase of the corresponding injected master pulse. The relative phases between these pulses are well defined as they are seeded by the same master pulse, however, collectively their global phase is random.

In order to prepare the slave pulses for interference to achieve the desired outputs, their relative phases need to be carefully controlled. This is achieved by manipulating the phase evolution of the master pulse which can be realised by introducing an amplitude perturbation to the electrical driving signal of the master laser \cite{Yuan2016} [Fig. \ref{scheme}b (i)]. This electrical modulation, with a temporal width of $\Delta t_m$, changes the carrier density in the laser cavity, which in turn alters the cavity refractive index and causes a temporary optical frequency shift of $\Delta \nu$, thus the photons produced after the modulation experience a phase shift of $\Delta \phi = 2\pi \Delta \nu \Delta t_m$ \cite{Yuan2016}. By locating the modulation in the interval between the onsets of two slave pulses, this phase difference can be transferred on to the slave pulses. As shown in Fig. \ref{scheme}b, the relative phases between the three slave pulses, $\phi_{12}$ and $\phi_{23}$, can be implemented independently by adding two small electrical perturbations to the master laser.

The prepared slave pulses then pass through an AMZI with one of its arms having a delay line that matches with the temporal separation of the slave pulses, resulting in interferences between consecutive slave pulses. As shown in Fig. \ref{scheme}b (iv), at the outputs of the AMZI, three pulses are formed within a single logical bit: two of them with their intensities and the relative phase completely determined by $\phi_{12}$ and $\phi_{23}$, whereas the third pulse has a random intensity due to the interference of two slave pulses originating from different master pulses with random phase relation (indicated in grey shading). As a result, the first two pulses could be used to represent the early and late bin for time-bin encoding.

\begin{figure*}[ht!]
  \centering
  \subfigure[ ]{\includegraphics[width=183.46pt,height=139.626pt]{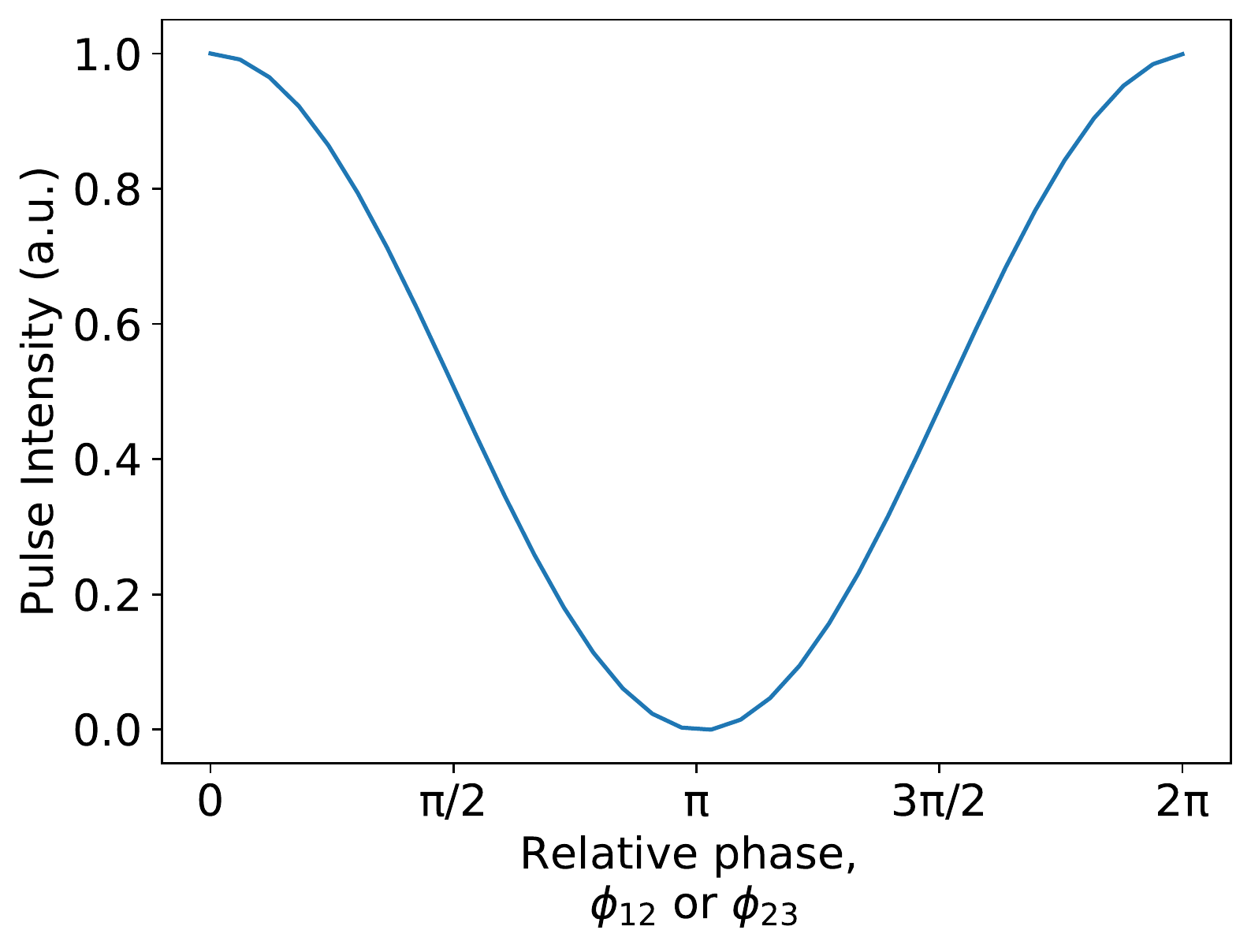}}\quad
  \subfigure[ ]{\includegraphics[height=140.6279pt,width=170.68767pt]{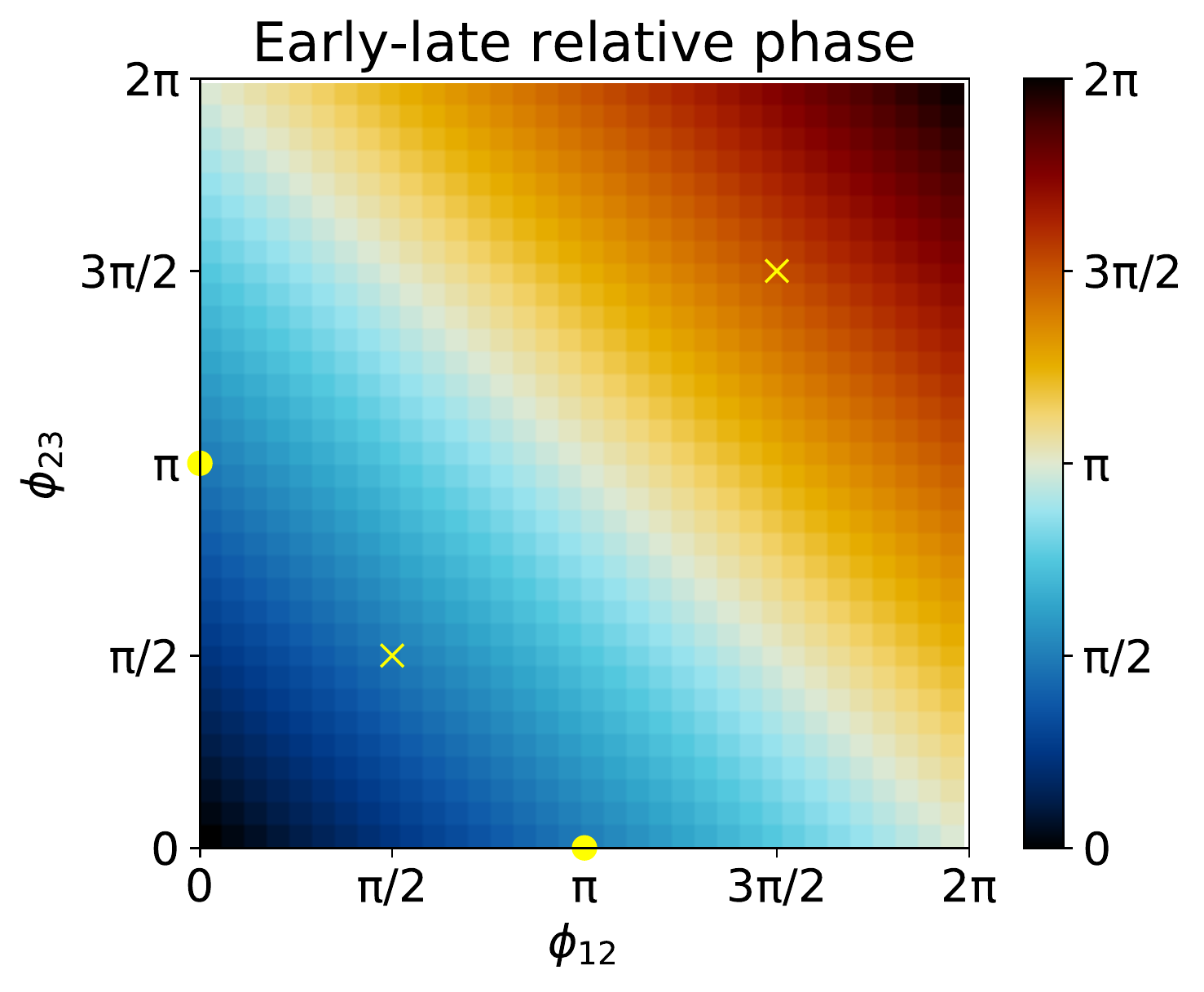}}
\caption{(a) Simulated intensity of the final output pulse as a function of the relative phase between the two slave pulses. For the pulse in the early (late) time bin, the intensity is determined by $\phi_{12}$ ($\phi_{23}$) (b) Simulated relative phase between the final output pulse pair, $\phi_{EL}$, as a function of $\phi_{12}$ and $\phi_{23}$. For Y-basis (Z-basis) encoding, the suitable values for $\phi_{12}$ and $\phi_{23}$ are marked by yellow crosses (dots).}
\label{simulation}
\end{figure*}

To express the relative phase between the early and late time bins and their intensities in terms of $\phi_{12}$ and $\phi_{23}$, we consider the pulses generated by the slave laser as three coherent states  $\hs \ketn{\alpha_1}$, $\hs \ketn{\alpha_2}$ and $\hs \ketn{\alpha_3}$, with amplitude \textit{A}:

\eqn{\hs \ketn{\alpha_1} \ee \ketn{Ae^{i(\omega t + \phi_1)}}\nn \\
\hs\ketn{\alpha_2} \ee \ketn{Ae^{i(\omega t + \phi_1 + \phi_{12})}} \nn \\
\hs\ketn{\alpha_3} \ee \ketn{Ae^{i(\omega t + \phi_1 + \phi_{12} + \phi_{23})}}
\label{waves}}
where the phase of the first coherent state, $\phi_1$, is uniformly distributed over $[0,2\pi)$.

\noindent In the AMZI, the interference between $\hs \ketn{\alpha_1}$ and $\hs\ketn{\alpha_2}$ ($\hs \ketn{\alpha_2}$ and $\hs \ketn{\alpha_3}$) gives rise to the early (late) time bin $\hs \ketn{\alpha_E}$ ($\hs \ketn{\alpha_L}$), which can be expressed as:

\eqn{\ketn{\alpha_E} \ee  \frac{A}{2} e^{i \left(\omega t  +\phi_1\right)} \left(1+e^{i \phi_{12}}\right)  \nn \\
\ketn{\alpha_L} \ee \frac{A}{2} e^{i \left(\omega t  +\phi_1 + \phi_{12} \right)} \left(1+e^{i \phi_{23}}\right)    \label{EL}}

\noindent and their corresponding intensities and phases are given by

\eqn{r_E \ee A\cos\bk{\frac{\phi_{12}}{2}}, \hs \phi_E = \omega t +\phi_1+\frac{\phi_{12}}{2} \nn\\
r_L \ee A\cos\bk{\frac{\phi_{23}}{2}}, \hs \phi_L = \omega t +\phi_1+\phi_{12}+\frac{\phi_{23}}{2}  \label{rR0}}

\noindent respectively. The relative phase between the early and late time bins $\phi_{EL}$ and their intensities are simulated based on Eqn. \ref{rR0} and shown in Fig. \ref{simulation}. This scheme could therefore be applied to time-bin based BB84 decoy-state QKD with Z and Y basis encoding. For Z-basis encoding, a pulse is located in either the early time bin (representing bit 0) or the late time bin (representing bit 1). To encode bit 0, $\phi_{12}$ is set to 0 to produce a pulse with maximum intensity in the early time bin and $\phi_{23}$ is set to $\pi$ to suppress any light in the late time bin. Similarly, bit 1 can be encoded by choosing $\phi_{12}$ = $\pi$ and $\phi_{23}$ = 0.

A decoy state in the Z basis can be generated in a similar way as described above. Instead of using zero relative phase which results in a pulse with maximum intensity, a decoy state with a lower intensity can be generated by choosing a relative phase close to $\pi$, according to Fig. \ref{simulation}a. For example, a decoy bit-0 state with an intensity of 0.1 can be generated by choosing  $\phi_{12}$ = 0.9$\pi$ and $\phi_{23}$ = $\pi$. Therefore, the flexibility to adjust the intensity level of the decoy state is enabled simply by implementing the appropriate relative phases, which itself is controlled by simple modulation of the electrical drive signal applied to the master laser.

\begin{figure}[tbph]

	\includegraphics[width=243.083pt, height=184.17pt]{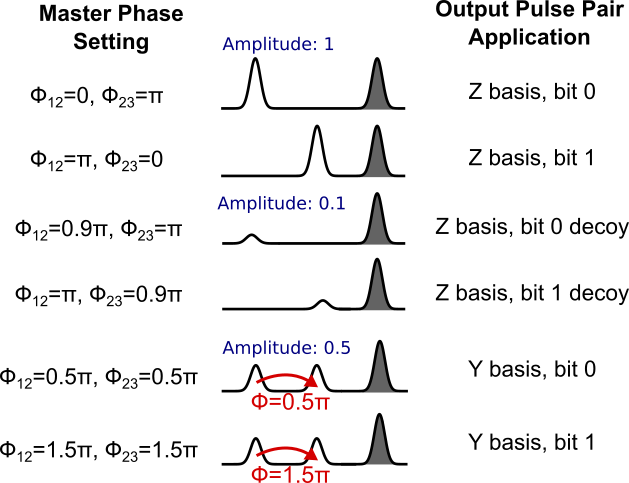}
	\caption{Direct generation of possible QKD states with the corresponding phase settings on the master laser. }
	\label{encoding}
\end{figure}

In the Y-basis, a single bit comprises both the early and late time bins with a relative phase of $\pi/2$ (bit 0) or $3\pi/2$ (bit 1). Each time bin has half the intensity of the signal state in the Z basis. From Eqn. \ref{EL}, the relative phase between the early and late time bins is simply $\phi_{EL}$ = ($\phi_{12}$  + $\phi_{23}$)/2. Since the intensities of the early and late time bins must be equal, it is necessary that $\phi_{12}$ = $\phi_{23}$. As a result, to encode bit 0 with $\phi_{EL}$ = $\pi$/2, $\phi_{12}$ = $\phi_{23}$ = $\pi$/2. Similarly,  to encode bit 1 with $\phi_{EL}$ = $3\pi$/2, $\phi_{12}$ = $\phi_{23}$ = 3$\pi$/2. A summary of the phase settings for various potential encoding states including an example of decoy state is illustrated in Fig.~\ref{encoding}.

\section{Results}
\begin{figure}[tbph!]

    \includegraphics[width=205.43958pt,height=252.14pt]{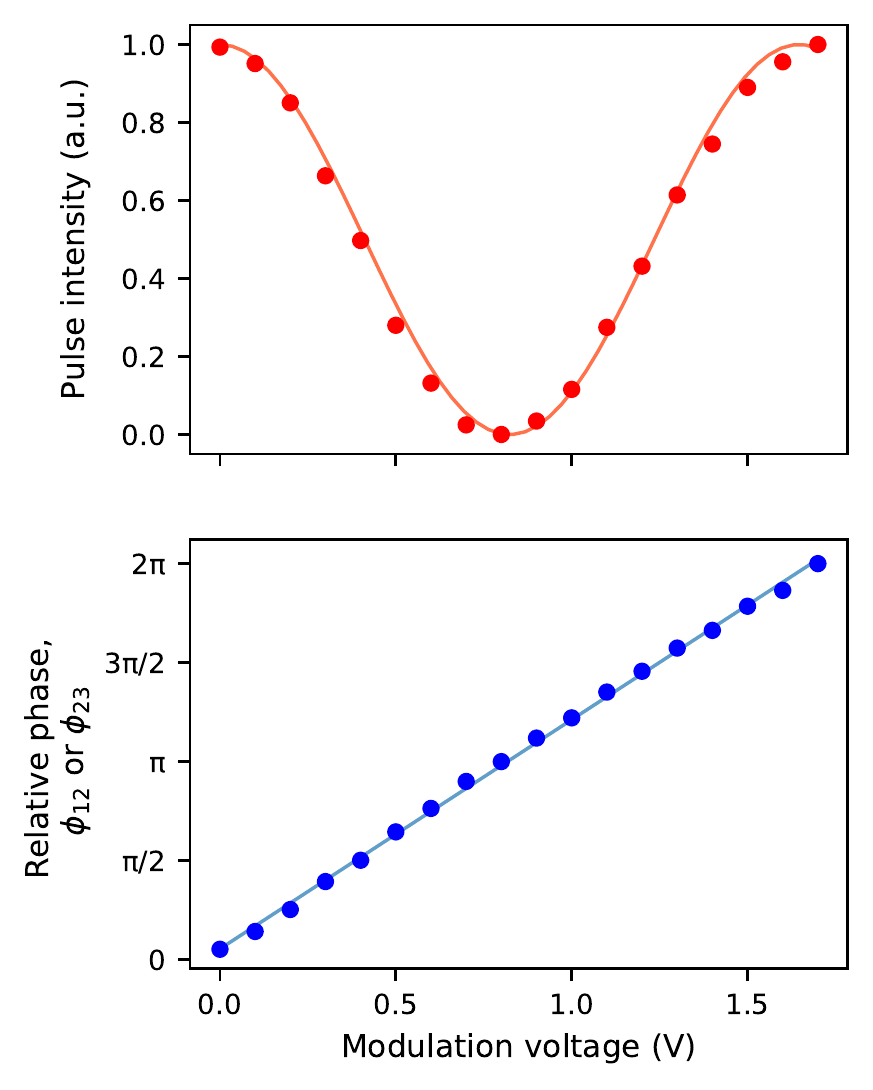}

	\caption{Experimental characterisation of the pulse generated at the outputs of the AMZI. Intensity of an individual pulse (top) and the relative phase between a pulse pair (bottom) as a function of modulation voltage applied to the master laser. A cosine (linear) fit is applied to the top (bottom) plot. }
	\label{phaseshift}
\end{figure}

A key element to implement our proposed scheme is precise control of the relative phases between slave laser pulses, $\phi_{12}$ and $\phi_{23}$, as they completely determine the final output states. This can be achieved by carefully adjusting the amplitude of the modulation applied to the master laser's electrical signal. The master laser is operated at 667 MHz and the slave laser at 2 GHz so that every master pulse is long enough to seed three slave pulses. A modulation with a fixed temporal width of 150 ps is applied to the electrical signal between the onsets of two slave pulses and its voltage amplitude is varied. The amplitude of the pulse at the output of the AMZI is measured as a function of modulation voltage, as shown in Fig. \ref{phaseshift}a, confirming the ability to continuously tune the transmitter output pulse intensity. Fig. \ref{phaseshift}b shows that the half-wave voltage, V$_{\pi}$ is around 0.8 V, which is significantly lower than that of common LiNbO$_3$ phase modulators. The minor deviation from the theoretical values can be attributed to the imperfections in experimental equipment (e.g. phase noise in lasers).

\begin{figure*}[tbph!]

	\includegraphics[width=446.33pt,height=249.32625pt]{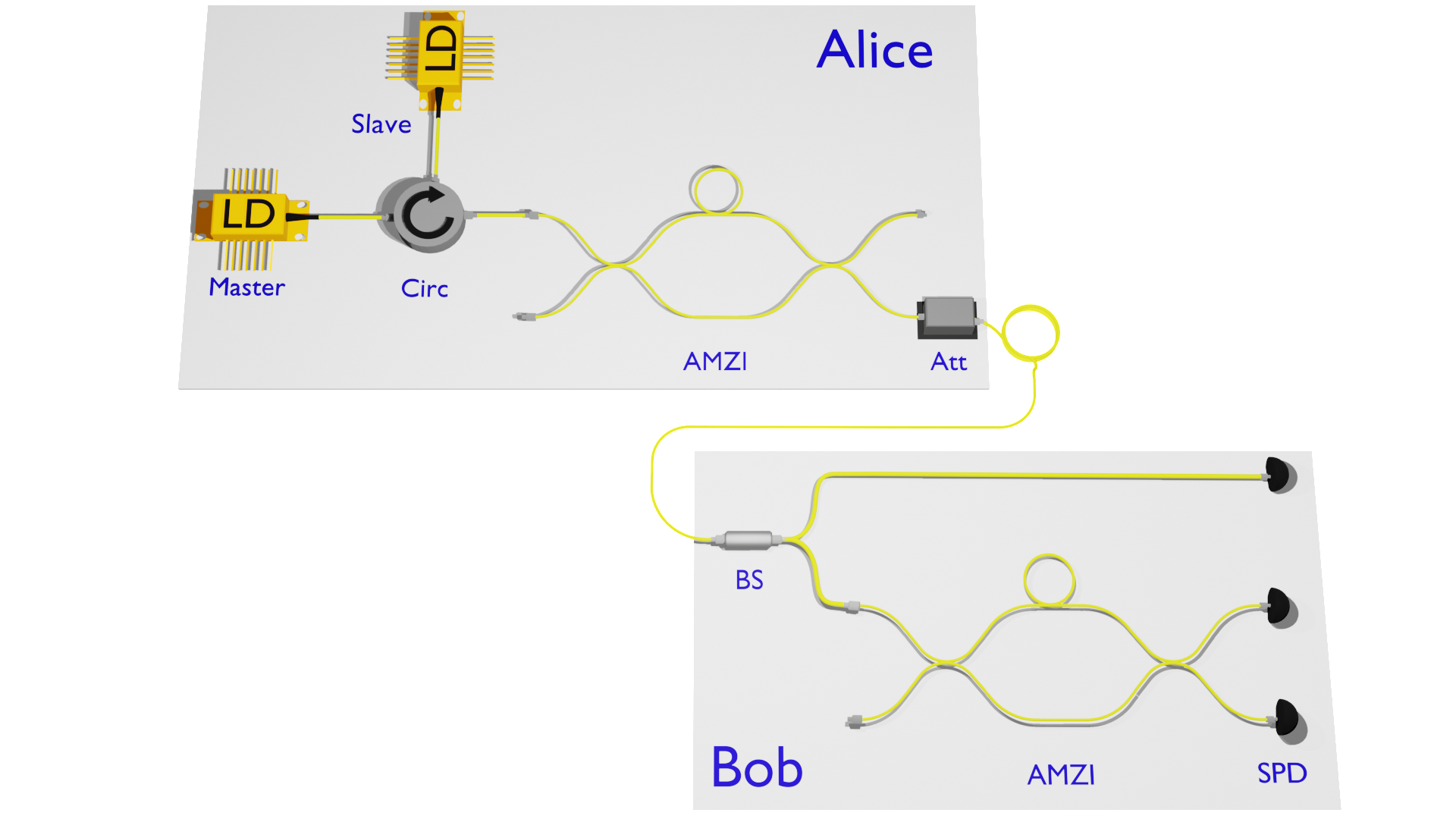}
	\caption{Experimental setup for BB84 protocol. LD, laser diode; Circ, circulator; AMZI, asymmetric Mach-Zehnder
interferometer; Att, attenuator; BS, beamsplitter; SPD, single-photon detector.}
	\label{setupphasedecoy}
\end{figure*}

\begin{figure*}[tbph!]
    
	\includegraphics[width=432.15942pt,height=201.48204pt ]{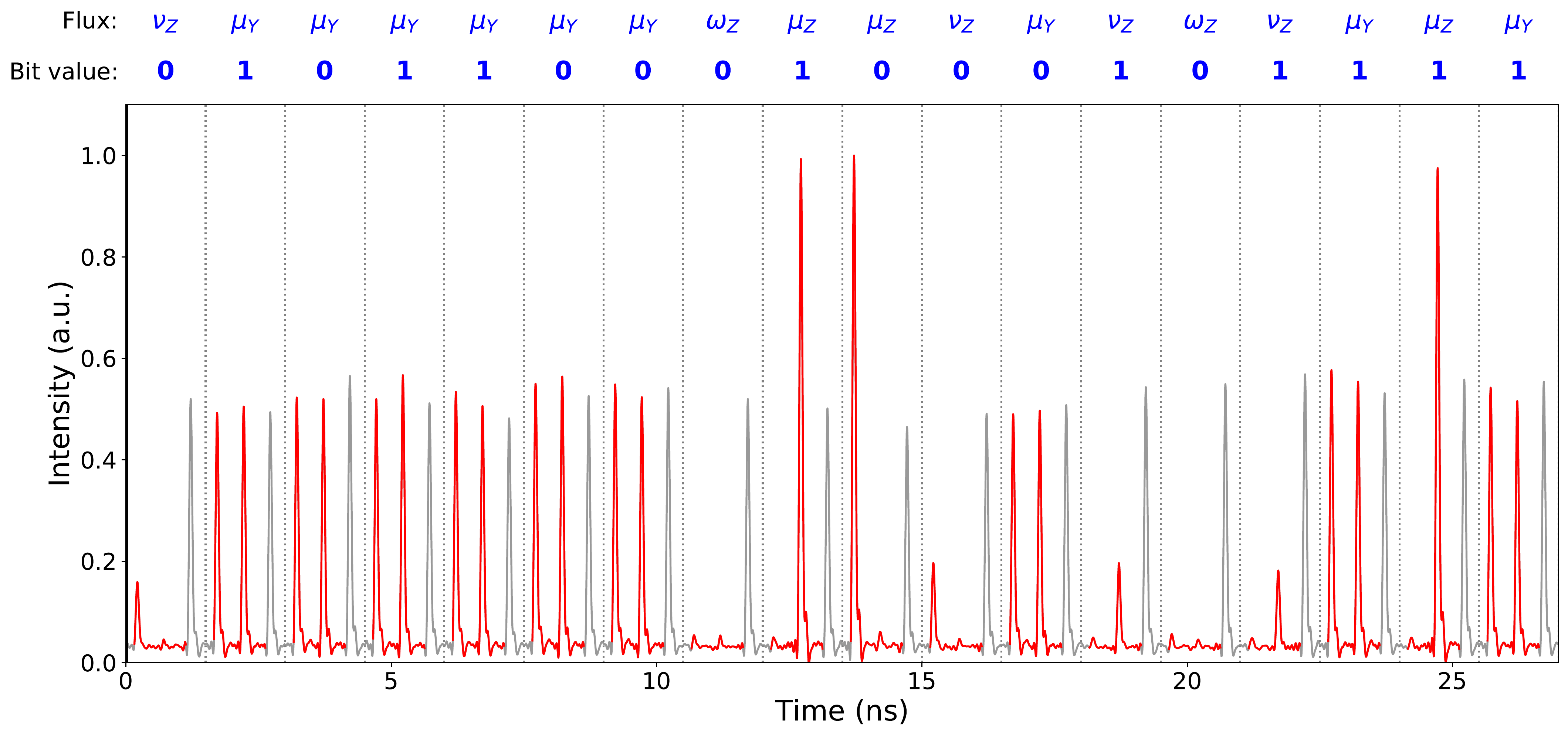}
	\caption{Time-averaged pulse pattern generated with direct modulation scheme. The corresponding flux and bit value for each bit are indicated at the top of the figure. The red pulses are used for QKD operation whereas the grey pulses are phase-randomised pulses. Signal states ($\mu$) and decoy states ($\nu$ and $\omega$) can be readily generated without any external modulator.}
	\label{pulsetrain}
\end{figure*}

To demonstrate the potential of our scheme for QKD, we implement a BB84 protocol with two decoy
states \cite{Ma2005}.
The experimental setup is shown in Fig. \ref{setupphasedecoy}.
The outputs of Alice (transmitter) (Fig. \ref{pulsetrain}) consist of a random mixture of the signal states with intensity $\mu$ prepared in the Z and Y bases and the decoy states with intensities $\nu$ and $\omega$ prepared in the Z basis, where $\mu > \nu > \omega$. 
The intensity levels of the decoy states can be accurately adjusted to maximise the key rate performance. A variable optical attenuator is placed before the output of Alice in order to attenuate the signals to the desired mean photon number level. Bob (receiver) adopts passive basis choice using a beamsplitter. In the Z basis, the photons are directly detected by a single-photon detector (SPD) where the bit value can be retrieved from their arrival time using a time-tagger. In the Y basis, the photons pass through an AMZI which results in three interfering pulses within a bit. Only the first interfering pulse is measured as it is originated from the interference between the early and late time bins. The phase basis of the AMZI is adjusted such that bits 0 and 1 correspond to the detections in different detectors. The other two interfering pulses involve the interference of photons with no deterministic phase difference and they are not processed to use for key generation (similar to the traditional processing scheme for detecting phase-encoded time bins using an AMZI at Bob~\cite{Yuan2016}). Very slight variation in pulse heights in Fig. \ref{pulsetrain} is related to the finite bandwidth of real-world high-speed components. This has been observed in other QKD transmitter designs too, not related to our new approach introduced here. The study of such real-world encoding imperfections is a topic in its own right and various solutions have been proposed including variations to the security proofs and post-processing \cite{Sixto2022, Yoshino2018}.

\begin{figure}[tbph!]
	\includegraphics[height=196.7679pt,width=223.9423pt]{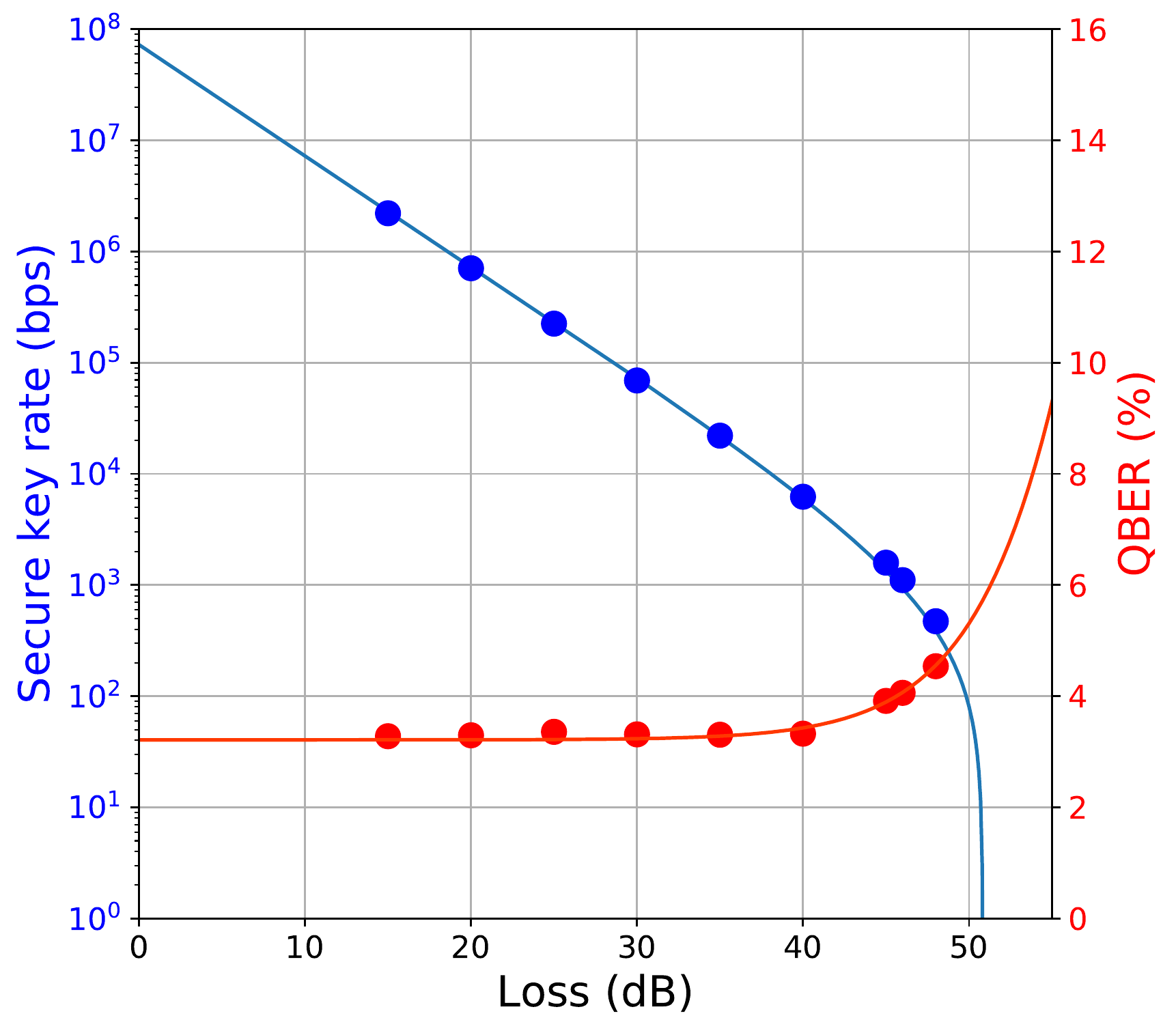}
	\caption{Key rates and QBER performance of the BB84 protocol carried out by our directly modulated transmitter scheme. The experimental data (dots) are consistent with simulated rates (lines). }
	\label{skr}
\end{figure}

In our proof-of-principle QKD experiment, we implement a standard, asymptotic, decoy-state BB84 analysis \cite{Ma2005} which does not explicitly consider the presence of the extra pulses inherent to our modulation scheme. A full security proof is beyond the scope of this work, but we provide some arguments as to why this should not represent an issue in the discussion section. The quantum bit error rate (QBER) is measured and used to compute the secure key rate (SKR), as shown in Fig. \ref{skr}. Positive key rates can extend up to a channel loss of 48 dB (equivalent to 240 km of standard fiber with an attenuation of 0.2 dB/km). A secure key rate of 2.21 Mbps is measured at 15 dB (75 km), demonstrating the suitability of our system for metro-scale QKD networks. The QBER can be maintained at a base level of 3.3\% before the detector noise becomes comparable to the signal counts at high channel losses. This is comparable to the performance achieved by QKD systems using conventional phase and intensity modulators \cite{Lucamarini2013, Yuan2018}.

\section{Discussion}

We have demonstrated a simple scheme to generate phase- and intensity-tunable pulses at GHz clock speeds, which can implement the BB84 protocol without the need for any phase or intensity modulators. As shown above, the performance of QKD based on our scheme approaches that of conventional LiNbO$_3$ modulators. We attribute this feature to the adoption of OIL which reduces the timing jitter and the frequency chirp in the output pulse significantly \cite{Lau2009, Yuan2014}, whilst maintaining a coherent phase transfer from the master to the slave laser.

The presence of the additional pulses in this modulation method means it is not completely trivial to apply the security proof for a standard scheme \cite{Ma2005}. The concern would be that Eve could somehow break the security by attacking these extra pulses. However, this is unlikely to be true as the state in these extra time bins is essentially obfuscated by the phase randomisation procedure. Additionally, well known uncertainty relations between phase and photon number further constrain Eve's ability to extract relevant information. In Appendix A we describe these arguments in more detail and provide a sketch for how a fully general security proof could be carried out.

Compared to the common approach where dedicated phase and intensity modulators are required in the transmitter to generate the encoding states and the decoy states, our scheme allows all such states to be generated directly from two lasers and an AMZI by exploiting direct phase modulation technique \cite{Yuan2016} and coherent interference. In this way, not only do we remove the modulators but also the high-speed RF signals and power supplies necessary to drive the modulators, thereby reducing the complexity and the cost of a QKD system significantly.

As our transmitter only has two active components (i.e. the lasers), the power consumption is expected to be low. Together with the low V$_{\pi}$, the design is well-suited for on-chip integration \cite{Paraiso2019}, offering a route to compact, low cost and power efficient quantum transmitters.
Beyond QKD, this simple approach to generating intensity- and phase-variable pulses could find other applications in classical optical communications, where the ability to precisely manipulate intensity and phase enables novel high-density encoding schemes for pushing communication bit rates.

In conclusion, we have demonstrated a scheme to directly generate phase- and intensity-tunable pulses at high speed, using two gain-switching lasers in an OIL configuration with an AMZI.
By applying appropriate electrical driving signals to the lasers, the intensity and phase of the pulses can be simply varied.
The design is shown to have strong potential as a QKD transmitter for decoy-state QKD, where all required encoding and decoy states for a BB84 protocol can be directly generated without any bulk modulators. 
Thus, our scheme offer a new possibility to perform QKD using compact, low-cost, yet high-performance devices, advancing the development of quantum communications towards larger scale deployments.

\section{Appendix A}
\subsection{Experimental setup}
The transmitter consists of two independent DFB lasers with a 10 GHz modulation bandwidth and an integrated thermoelectric cooler, operating at 1550 nm. The two lasers are connected through a circulator, allowing light to be injected from the master to the slave laser. A variable optical attenuator is used to adjust the injection power. Each laser is driven by an arbitrary waveform generator with a sampling rate of 24 GS/s and 10-bit vertical resolution. The RF driving signal is amplified by an RF amplifier and then combined with a DC bias via a bias-tee. The two modulations on the master RF signals have a temporal width of 150 ps and a separation of 450 ps from each other. The modulation level depends on the desired outputs. The delay between the RF signals of the two lasers are temporally aligned with picosecond resolution to ensure that the slave pulses are coherently seeded by the correct master pulses. The master (slave) laser is driven at a clock rate of 667 MHz with a on-time of 1.4 ns (2 GHz with an on-time of 300 ps). The AMZIs placed in the transmitter and the receiver are chip-based interferometers. Each of them has a delay line of 500 ps and an integrated heater which can be controlled electronically in one of its arms. The heater acts as a phase shifter which is used to tune the phase delay between the two arms and align the phase basis between the transmitter and the receiver. An optical filter is also used in the transmitter to reduce the noise and enhance the phase coherence. The channel loss is emulated using a variable optical attenuator. A superconducting nanowire single photon detector with $\sim$70\% efficiency and 50 Hz dark counts is used in the receiver. The detection events are measured with a 100 ps resolution time tagger. A central window of 300 ps of the time bin is selected in order to suppress the errors due to timing jitter. 

\subsection{QKD protocol}
We implement the two-decoy-state BB84 protocol in the asymptotic case \cite{Ma2005} with imbalanced basis choice where the Y (Z) basis is selected with a probability of 90\% (10\%), i.e. Y (Z) basis is the majority (minority) basis. The average photon numbers of the signal ($\mu$), decoy ($\nu$) and vacuum ($\omega$) states are 0.4, 0.16 and 0.015, respectively. Alice randomly prepares $\mu$, $\nu$ and $\omega$ in the Z basis but only prepare the $\mu$ in the Y basis. To match with Alice's basis-sending probability, Bob uses a beamsplitter with a splitting ratio of 90:10 to implement passive basis choice where Z basis is chosen with a probability 10\% and Y basis is chosen with a probability of 90\%. The key bits are extracted from the Y basis only whereas the Z basis is used to estimate the information leakage. The gain and QBER for each state are measured to estimate the final secure key rate analytically \cite{Ma2005}.

\subsection{Security discussion \label{sketch}}
Here we provide some more details about the additional security considerations that may arise due to the additional pulse that arises in our modulation scheme and sketch how the standard decoy-state BB84 security proof could be modified to account for these. The two issues to keep in mind are i) whether any information about the encoded bits is leaked directly or ii) whether the global phase randomisation (and hence the decoy-state analysis) is compromised, potentially overestimating the secret key rate.

\subsubsection{Modulation scheme}
\begin{figure}[htbp]

\includegraphics[height=54.6187pt,width=241.0789pt]{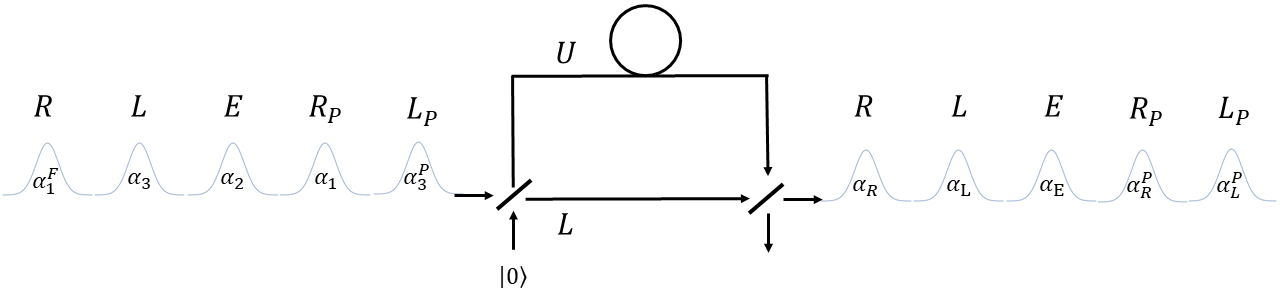}
\caption{Schematic of direct phase and intensity modulation technique. The incoming pulse train of five time bins shows a complete triplet $\alpha_{1,2,3}$ along with the first pulse of the following triplet $\alpha_{1}^F$ and the last pulse of the preceding triplet $\alpha_{3}^P$. A delay in the top arm of the AMZI causes interference between successive pulses, which facilitates modulation of the relative phase and intensity of the output state in the early ($E$) and late ($L$) time bins.} 
\label{pulse}

\end{figure}

We begin by describing the modulation scheme in Fig.~\ref{scheme} in more detail. Each encoding is created from an initial triplet of pulses ($\ket{\alpha_1}$, $\ket{\alpha_2}$, $\ket{\alpha_3}$) passed through an AMZI, leading to an output triplet ($\ket{\alpha_1}$, $\ket{\alpha_2}$, $\ket{\alpha_3}$) where the key is encoded in the phase difference between the pulses in the first two time bins, labelled early (E) and late (L), followed by a unused, randomised pulse in the so-called random (R) bin. To fully capture all the potentially relevant correlations we need to also consider the pulses either side of a given encoding (i.e.\ last pulse of the preceding triplet, $\ket{\alpha_3^P}$, and the first pulse of the following triplet, $\ket{\alpha_1^F}$). The action of the AMZI is to mix each pulse with a vacuum state at the input beamsplitter and then delay the upper (U) arm to be recombined with the subsequent pulse at the final beamsplitter. This means that, before the AMZI, the first pulse of a given encoding triplet occupies the time bin associated with the random pulse of the preceding triplet. In other words the states in the various time bins before the AMZI are given by (see also Fig.~\ref{pulse})

\begin{align*}
L_P: \hs \ketn{\alpha_3^P} &= \ketn{Ae^{i(\omega t + \phi_1 + \phi_R^P)}}\nn \\
R_P: \hs \ketn{\alpha_1} &=\ketn{Ae^{i(\omega t + \phi_1)}}\nn \\
E: \hs\ketn{\alpha_2}  &=\ketn{Ae^{i(\omega t + \phi_1 + \phi_{12})}} \nn \\
L: \hs\ketn{\alpha_3}  &=\ketn{Ae^{i(\omega t + \phi_1 + \phi_{12} + \phi_{23})}}\nn \\
R: \hs\ketn{\alpha_1^F}  &=\ketn{Ae^{i(\omega t + \phi_1 + \phi_{12}+\phi_{23} + \phi_R^F)}}\tag{A1}
\end{align*}

where $A\in \mathbb{R}$ is the input intensity of each pulse and $\phi_{12}$ and $\phi_{23} $ are the relative phases that are chosen to encode one of the four BB84 states. Note that $\phi_1$ must be uniformly distributed over $[0,2\pi)$ in order to make the output ensemble phase randomised. The preceding and following triplets also have a randomised phase, which for brevity we here write relative to the first triplets phases via the variables $\phi_R^F, \phi_R^P$, which is therefore also uniformly distributed over  $[0,2\pi)$.

These states can be propagated through the initial beamsplitter, time delay and final beamsplitter to derive the following output states,

\begin{align*}
\ketn{\alpha_{R_P}} &=  \frac{A}{2} \left(1+e^{i \phi^P_{R}}\right) e^{i \left(\omega t  +\phi_1\right)} \nn \\
\ketn{\alpha_E} &=  \frac{A}{2} e^{i \left(\omega t  +\phi_1\right)} \left(1+e^{i \phi_{12}}\right)  \nn \\
\ketn{\alpha_L} &= \frac{A}{2} e^{i \left(\omega t  +\phi_1 + \phi_{12} \right)} \left(1+e^{i \phi_{23}}\right)   \nn\\
\ketn{\alpha_R} &=   \frac{A}{2} e^{i \left(\omega t  +\phi_1 + \phi_{12} + \phi_{23} \right)} \left(1+e^{i \phi_{R}^F}\right) 
\label{ER}
\tag{A2}
\end{align*}

The complex amplitude describing a coherent state can be expressed as a phase and an intensity, $\ket{\alpha} = \ket{r e^{i\phi}}$, where

\begin{align*}
r &= \nn \sqrt{\mathrm{Re}(\alpha)^2 + \mathrm{Im}(\alpha)^2}  \\
\phi &= \tan^{-1}\bk{\frac{\mathrm{Im}(\alpha)}{\mathrm{Re}(\alpha)}}
\tag{A3}
\end{align*}

which gives,

\begin{align*}
r_{R_P} &= A\cos\bk{\frac{\phi_R^P}{2}}, \hs \phi_{R_P} = \omega t+ \frac{\phi_R^P}{2} +\phi_1\nn \\
r_E &= A\cos\bk{\frac{\phi_{12}}{2}}, \hs \phi_E = \omega t +\phi_1+\frac{\phi_{12}}{2} \nn\\
r_L &= A\cos\bk{\frac{\phi_{23}}{2}}, \hs \phi_L = \omega t +\phi_1+\phi_{12}+\frac{\phi_{23}}{2}  \label{rR}\\
r_R &= A\cos\bk{\frac{\phi_R^F}{2}}, \hs \phi_R = \omega t+ \frac{\phi_R^F}{2} +\phi_1+\phi_{12}+\phi_{23} \nn
\tag{A4}
\end{align*}

From this we can immediately verify the claims in the main text that the intensity of the $E$ and $L$ bins are controlled by $\phi_{12}$ and $\phi_{23}$ and that the phase difference between the $E$ and $L$ bins is given by $\phi_{EL} = (\phi_{12}  + \phi_{23})/2$, which allows the following encoding pattern for all four BB84 states

\begin{align*}
\ket{0}:\phi_{12} &= 0, \phi_{23} = \pi, &  \ket{1}:\phi_{12} &=\pi, \phi_{23} = 0 \nn \\
\ket{+}:\phi_{12} &= \frac{\pi}{2}, \phi_{23} =\frac{\pi}{2}, &  \ket{-}:\phi_{12} &=\frac{3\pi}{2}, \phi_{23} =\frac{3\pi}{2}  \label{bb84phi}
\tag{A5}
\end{align*}

\subsubsection{Security considerations}
Turning to the security implications of the the `random' pulses, straightforward substitution of Eq.~\ref{bb84phi} into Eq.~\ref{ER} shows that the coherent amplitude of both the $R$ and $R_P$ bins is identical for all encoding choices (this is because for all settings in Eq.~\ref{bb84phi} it holds that $\phi_{12}+\phi_{23} = \pi \mod 2\pi$. This might seem initially sufficient to argue that there is no additional information leakage due to the extra pulses. However, one should also consider the possibility that Eve could choose to combine different pulses in her attack. For instance the relative phases between the early at late bins and their adjoining randomised pulses are,

\begin{align*}
\phi_{LR} &= \frac{\phi_R^F+ \phi_{23}}{2} \nn \\
\phi_{ER_P} &= \frac{\phi_R^P - \phi_{12}}{2} \label{philr} 
\tag{A6}
\end{align*}
By themselves, the values of $\phi_{LR}$ and $\phi_{ER_P}$ leak no information because the phases that determine the secret key, $\phi_{12}$ and $\phi_{23}$, are effectively one-time padded by the uniformly distributed variables, $\phi_R^P$ and $\phi_R^F$. However, by considering Eq.~\ref{rR} we can see information about $\phi_R^P$ and $\phi_R^F$ could in turn be obtained by measuring the intensity of the random pulses ($r_{R_P}$ and $r_R$). Nevertheless, security can still be maintained provided it is impossible for Eve to simultaneously learn the relative phases of any two pulses and the corresponding intensities, which is the case due to the conjugate nature of the number and phase operators. A measurement that perfectly revealed that photon number in either the $R$ or $R_P$ bins would totally randomise the phase.   Moreover, even if Eve chooses to maximise her information about relative phases in Eq.~\ref{rR} this will tell her nothing about the absolute phase of each encoding triplet ($\phi_1$, $\phi_1^P$ etc) since learning the difference between two uniformly random variables leaks no information about either variable. Thus the phase randomisation condition required for a decoy state analysis is not compromised.

Although a full security proof is beyond the scope of this work, we provide a sketch of how one could proceed. Firstly, one would adapt the standard decoy argument to show that, from Eve's perspective at the ensemble level, the experimental scheme is indistinguishable from a scheme in which Alice and Bob prepare true single photon qubits in the $E$ and $L$ bins along with extra coherent states in the random bins. Then, construct a complete entanglement based version of this modulation scheme, including a fictitious measurement on a suitably prepared entangled state that determines the randomised phase of each encoding triplet and projectively prepares the appropriate coherent state in each random bin. The total  system would then be described by a pure state $\ket{X_ABER_AR}$ where  $R$ describes the extra coherent pulses and $R_A$ is a Alice's register of the randomised phase values (these are never used in the protocol so there is no actual need for Alice to possess this register, it is only necessary that Eve does not posses it). Then, in the worst case one would simply assume that Eve is given the entire $R$ system and one would then bound Eve's conditional entropy about the Alice's key generation measurements, $S(Z_A|ER)$. Note that this approach means no extra monitoring of the random bins is required. The previous arguments regarding the impossibility of learning the key from the $R$ register could be made quantitative by utilising tools such as entropic uncertainty relations for phase and photon number (e.g. \cite{Hall:1993bo,Coles:2017ec}). Some of these results require an upper bound on the energy, but this is simply given by the maximum encoding amplitude, $A$. These would be combined with the standard security arguments for the information leaked through Eve's purification of the channel describing the measured time bins \cite{Ma2005} would be sufficient to determine the secret key rate.

\subsection*{Acknowledgments}
\begin{acknowledgments}
We thank D. E. Browne for helpful discussions. We acknowledge funding from the European Union Horizon 2020 Research and Innovation Programme (Grant 857156 “OPENQKD”). Y.S.L. acknowledges financial support from the EPSRC funded CDT in Delivering Quantum Technologies (EP/L015242/1) and Toshiba Europe Ltd.
\end{acknowledgments}

\subsection{DISCLOSURES}
The authors declare no conflicts of interest.

\bibliography{library_new}

\end{document}